# Adaptive video tranmsision using QUBO method and Digital Annealer based on Ising machine

Bo Wei*, Hang Song† and Jiro Katto*
*Department of Computer Science and Communication Engineering, Waseda University, Tokyo, Japan
†School of Engineering, The University of Tokyo, Tokyo, Japan
weibo@aoni.waseda.jp, songhang@g.ecc.u-tokyo.ac.jp, katto@waseda.jp

*Abstract*—With the dramatically increasing video streaming in the total network traffic, it is critical to develop effective algorithms to promote the content delivery service of high quality. Adaptive bitrate (ABR) control is the most essential technique which determines the proper bitrate to be chosen based on network conditions, thus realize high-quality video streaming. In this paper, a novel ABR strategy is proposed based on Ising machine by using the quadratic unconstrained binary optimization (QUBO) method and Digital Annealer (DA) for the first time. The proposed method is evaluated by simulation with the real-world measured throughput, and compared with other state-of-the-art methods. Experiment results show that the proposed QUBO-based method can outperform the existing methods, which demonstrating the superior of the proposed QUBO-based method.

*Keywords— DASH, adaptive bitrate control, Ising machine, QUBO, Digital Annealer*

## I. Introduction

With the explosion of COVID19, it has become normal to attend meeting, education or event virtually, which leads to the fast growing of video streaming. Video traffic has been the majority of the global IP traffic, which accounts for over 80% of the total internet traffic [1]. In addition, the 5th generation of mobile systems (5G) [2] has enabled high-resolution video transmission including 4K/8K video, even VR (Virtual Reality) and AR (Augmented Reality) delivery. To ensure high-quality video experience, the streaming should be adaptively controlled. The adaptive bitrate (ABR) control technique constitutes the main part to realize high quality of experience (QoE) for users.

Existing state-of-the-art ABR methods can be classified into three categories: Rate-based (RB) method, Buffer-based method [3], Hybrid methods. RB employs the throughput prediction [4-6] to determine the future bitrate selection. BB utilizes the current buffer state to choose the bitrate. Hybrid methods, such as learning-based method [7, 8] and control-theoretic [9, 10] method, use bandwidth prediction, buffer occupancy and other information to adaptively control the bitrate.

In this paper, we proposed an innovative method to deal with the adaptive video streaming problem using quadratic unconstrained binary optimization (QUBO) method. The purpose of this method is to convert the high-quality streaming into the QUBO problem which can be solved by quantum annealing or simulated annealing. In this research, the Digital Annealer (DA), a quantum-inspired hardware architecture [11], is utilized to minimize the derived QUBO. To the best of our knowledge, this proposal is the first ABR method in the world which utilizes QUBO and DA to solve the adaptive video streaming problem. Experiments were conducted to evaluate the proposed QUBO-ABR method and compare the performance with other ABR algorithms. Results indicated that the proposed QUBO-ABR method has the highest score in terms of QoE, which demonstrated the superior of the QUBO-ABR method.

The rest of this paper is organized as follows. The formulation of the proposed QUBO-ABR method is given in Section II. The experiment setting is shown in Section III. The results are presented in Section IV. Finally, the conclusion is made in Section V.

## II. QUBO-ABR Method Formulation

For general, the QUBO problem is formulated as follows:

$$f_Q(\boldsymbol{x}) = \sum \sum q_{ij} x_i x_j \qquad (1)$$

where $x_i$, $x_j$ are binary variables and $q_{ij}$ is the coefficient which determine the QUBO problem. Then, the problem is solved by finding a binary vector $\boldsymbol{x}$ which make $f_Q(\boldsymbol{x})$ have the minimal value. QUBO problem is also closely to the Ising machine which is generally defined as:

$$H(\sigma) = -\sum \sum J_{ij} \sigma_i \sigma_j - \sum h_i \sigma_i \qquad (2)$$

By introducing $\sigma = 2x - 1$ and $x = x^2$, (2) can be rewritten in the form of QUBO [12].

In this section, the design of the ABR method using QUBO method is presented. The purpose of the QUBO-ABR is to ensure high-QoE adaptive video streaming. Conventionally, the QoE includes the video quality, rebuffering and quality switch. To get a high QoE, the video should be selected as high quality. However, both rebuffering and the quality switch should stay in low level. Otherwise, the total QoE will be impaired.

For the video quality factor, the QUBO term is formulated as follows:

$$H_1 = -a \sum_{n=1}^{N} \left( \sum_{l=1}^{L} x_{n,l} \cdot q(l) \right) \qquad (3)$$

where $x_{n,l}$ is the selection of the *n*th segment and $q(l)$ is the quality of *l*th video level. Since the QUBO problem is to find the minimal value, this term is given the negative sign. The video quality is better when the QUBO function value is smaller.

For the quality switch factor, the QUBO term is formulated as follows:

$$H_2 = b \sum_{n=1}^{N} \left( \sum_{l=1}^{L} (x_{n,l} - x_{n-1,l}) \cdot q(l) \right)^2 \quad (4)$$

This term means to reduce the quality switch between two consecutive video segments. When the switch happens, this term will generate a positive value which gives a penalty to the total QUBO. Therefore, in the optimization, this term can ensure the quality switch in a low level.

For the rebuffering factor, the inequation is defined as follows:

$$\frac{\sum_{l=1}^{L} S(l) \cdot x_{n,l}}{C_{pred}} - B_n < 0 \quad (5)$$

where $B_n$ is the buffer state at the $n$th segment. $S(l)$ is the size of the segment to be download. $C_{pred}$ is the bandwidth prediction. This equation can ensure that the buffer length in the player is larger than the downloading time of the next segment. Therefore, the rebuffering event can be prevented. This equation can be transformed into QUBO using slack variables [12].

Finally, the one-hot constraint is applied to $x_{n,l}$ as (6) since for each segment, only one bitrate can be selected.

$$\sum_{n=1}^{N} \left( \sum_{l=1}^{L} x_{n,l} - 1 \right)^2 \quad (6)$$

III. EXPERIMENT AND RESULTS

*A. Experiment Setting*

In order to evaluate the performance of the proposed method and compare with existing ABR methods, simulation experiment was conducted by using the network traces measured in real world. The network condition of Tram and Ferry in the open dataset [13] were used as shown in Fig.1. And the QUBO problem is solved by using the DA provided by FUJITSU for research.

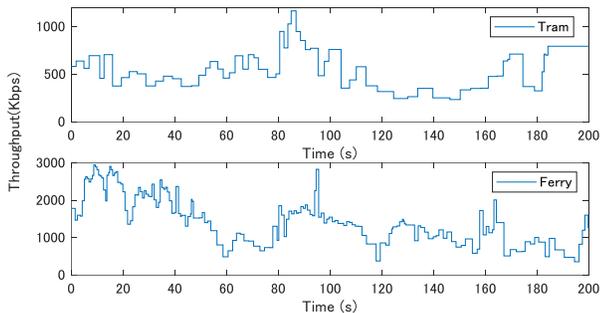

Fig. 1. Network traces for experiment.

*B. Evaluation Metrics*

In this work, the QoE definition is inherited from [9], which is calculated as:

$$QoE = \sum_{n=1}^{N} q[R_n] - w \sum_{n=1}^{N} T_{rebuf} - \sum_{n=1}^{N} |q[R_n] - q[R_{n-1}]| \quad (7)$$

where $q[R_n]$ is the quality of bitrate $R_n$. $T_{rebuf}$ is the rebuffering time. $w$ is the penalty coefficient applied to the rebuffering term.

IV. EXPERIMENT RESULTS

The experiment results are shown in TABLE I. It can be observed that the proposed QUBO-ABR method has the highest QoE compared with conventional methods. The buffer-based method BBA has the worst performance in the tram case. For the ferry case, the proposal also has the best performance compared with other methods.

TABLE I. QOE COMPARISON

| SCENARIO | ABR METHOD | | | |
|---|---|---|---|---|
| | QUBO | PENSIEVE | MPC | BBA |
| TRAM | 13.65 | 8.01 | 3.78 | -16.82 |
| FERRY | 50.8 | 48.8 | 36.55 | 37.65 |

From the experiment results, it can be found that the proposed QUBO-ABR method outperforms the other methods, demonstrating the superior of the proposed QUBO-based adaptive bitrate control method.

V. CONCLUSION

In this paper, an ABR method based on Ising machine with QUBO is proposed for high-QoE DASH streaming for the first time. This method utilizes QUBO to decide the selected bitrate and solve by a quantum-inspired hardware, Digital Annealer. The proposal is evaluated in two different network scenarios and compared with conventional ABR methods. Results demonstrate that the proposed QUBO-ABR method has the best performance with highest QoE. In the future, the propose method will be tested in various network conditions and continuously optimized.